\documentclass[twoside,10pt,a4paper]{newFNLstyle}

\usepackage{graphics}
\usepackage{cite}
\usepackage{amssymb}
\usepackage{graphicx}

\begin{document}

\volnumpagesyear{0}{0}{000--000}{2001}
\dates{received date}{revised date}{accepted date}

\title{NEW INSIGHT INTO THE FLUCTUATIONS OF THE MOVING VORTEX LATTICE: NON-GAUSSIAN NOISE AND L\'{E}VY FLIGHTS}

\authorsone{J.~Scola\thanks{Present address: Service de Physique de l'\'{E}tat Condens\'{e}, CNRS URA 2464, DSM/DRECAM/SPEC, CEA Saclay, 91191 Gif sur Yvette Cedex, France. joseph.scola@cea.fr},
A.~Pautrat, C.~Goupil, Ch.~Simon}
\affiliationone{CRISMAT/ENSICAEN et Universit\'{e} de Caen, CNRS UMR 6508, 14050 Caen, France,}

\authorstwo{B.~Domeng\`{e}s}
\affiliationtwo{LAMIP, Phillips/ENSICAEN, 14050 Caen, France}

\authorsthree{C.~Villard}
\affiliationthree{CRTBT/CRETA, CNRS UPR 5001, 38042 Grenoble, France}

\maketitle

\markboth{Instructions for Typesetting Camera-Ready Manuscripts}{Surname1, Surname2 and Surname3}



\begin{abstract}
We report measurements and analysis of the voltage noise due the to
vortex motion, performed in superconducting Niobium
micro-bridges.
Noise in such small systems exhibits important changes from the behavior commonly reported in macroscopic samples.
In the low biasing current regime, the voltage
fluctuations are shown to deviate substantially from the Gaussian
behavior which is systematically observed at higher currents in the so
called flux-flow regime. The responsibility of the spatial
inhomogeneities of the critical current in this deviation from
Gaussian behavior is emphasized. We also report on the first
investigation of the effect of an artificial pinning array on the
voltage noise statistics. It is shown that the fluctuations can lose
their stationarity, and exhibit a L\'{e}vy flight-like behavior.
\end{abstract}

\section{On vortex noise}

Dynamic properties of a current-driven vortex lattice are
determined by the competition between two antagonist mechanisms:
on the one hand, the random pinning potential, responsible for the
critical current, tends to distort the vortex lattice. But on the
other hand, the long-range magnetic interactions between vortices
are responsible for their collective behavior and tend to
stabilize a lattice. The way these two interactions balance each
other is still under debate, and is the subject of numerous
theoretical predictions, not only for vortex lattices, but also in
the larger field of disordered elastic systems \cite{elastic}.
Experimentally, the observation of Bragg peaks gives evidence for
the persistence of a long range order during vortex motion
(the flux-flow regime) \cite{simon71,pautrat05}. However, this
moving lattice can not be completely perfect because it keeps on
interacting with its pinning potential, as evidenced by the
persistence of a main critical current $I_c$, even at high biasing
current (e.g. \cite{placais94,campbell72}). These dynamic
interactions arise in very small fluctuations of the lattice
velocity ($\delta \mathbf{v}_L$) and of the magnetic field density
($\propto \delta \mathbf{B}$). The measurement and analysis of
these fluctuations can give important information on the
underlying static and dynamic features.

Experimentally, vortex dynamics can be easily probed by
transport measurements: the flow of the current-driven vortex
lattice generates an electric field in the sample ($\mathbf{E}=
-\mathbf{v}_L \times \mathbf{B}$). The fluctuations induced by the
pinning interactions are reflected into the measurable voltage
drop across the sample:
\begin{displaymath}
\label{josephsondifferenciee}
\delta V (t) = - \int \left( \delta \mathbf{v}_L (t) \times \overline{\mathbf{B}}
    + \delta \mathbf{B}(t) \times \overline{\mathbf{v}_L} \right) \cdot \mathbf{dl},
\end{displaymath}
where the overlined symbols stand for the mean values. Since the
$\delta V(t)$ signal results from a random mechanism, it is
processed as a noise. Most of the studies carried out on vortex
noise have focused on the power spectral density (PSD), which
gives partial information on the phenomenon. In order to improve
the description of the noise process, the analysis of its power
distribution is sometimes necessary. It is generally assumed to be Gaussian \cite{clem}, as it is expected for
independent fluctuators due to the general applicability of the
Central Limit Theorem (CLT). It is interesting to notice the striking exception
that takes place in the low-$T_c$ superconductor $2H$-NbSe$_2$ in
the vicinity of the peak effect (i.e. an anomalous increase of
$I_c$ at high field). Two macroscopic vortex states with different
pinning properties have been observed to coexist in this regime
\cite{marchevsky01,pautrat05}. At the same time, the noise is
unconventional, marked by an extremely high level, and exhibits
strong non-Gaussian features \cite{marley95,rabin98}. These
combined effects have been explained using a model of random
creation and annihilation of metastable vortex states
\cite{paltiel}. To our knowledge, this peculiar case, where macroscopic
inhomogeneities are present, is the sole direct measurement of a clear deviation
from Gaussian fluctuations in the moving vortex state. Since vortex noise is most
generally recorded in large systems (typically centimetric), statistical averaging can make the deviation very difficult to observe.
Here, we study small bridges where fluctuations are expected to be fewer.
In such small systems, non-Gaussian effects are more likely observable \cite{mike}.

\begin{figure}
\centering{\resizebox{6cm}{!}{\includegraphics{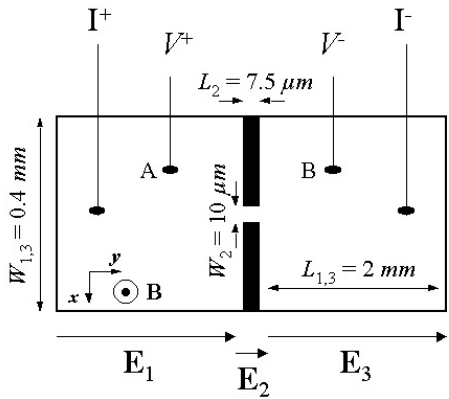}}}
\caption{Experimental geometry: the current flows along the $y$ axis, while the vortices flow along the $x$ axis.
The constriction (heavy black lines) magnifies the current density and lowers the critical current so that the lattice within the constriction depins for smaller current. In our experimental conditions, only the vortices in the constriction (so-called micro-bridge) move, the rest of the lattice being immobile and noiseless.
The grooves are etched at the center of the constriction, along the $y$ axis.}
\label{schemamp}
\end{figure}

The experiments were performed on micro-bridges made on a
$450nm$-thick film of Nb, which was ion-beam deposited on a sapphire
substrate. The pinning properties and
superconducting parameters of similar samples are given in ref.
\cite{pautrat04}. The films present a relatively smooth surface
(rms roughness is less than $5nm$ at the scale of few microns). In order to create
an artificial pinning potential for the vortices, twelve
rectangular grooves ($12\mu m \times 30nm$, $30nm$ deep) regularly spaced by
$300nm$ have been etched at the surface of one bridge, using a
focused ion beam. In our experimental configuration, the
grooves are parallel to the current direction, i.e.
perpendicular to the vortex velocity direction (as sketched in fig. \ref{BruitdeI}). The magnetic field
which creates the vortex lattice is applied perpendicular to the
film surface. The experimental geometry is summarized in
the figure \ref{schemamp}.

One of the aims of the experiment was to compare the vortex noise
properties between the regular micro-bridge and the micro-bridge
with the artificial surface state. For this purpose, we have
performed a systematic analysis of the voltage noise, with a
particular care on its statistics. The voltage time series were
measured with the four probes method. The detailed set-up can be
found in the ref. \cite{scola05}. The raw time signal was divided
into segments of $\Delta t = 0.3s$, each of them being
Fourier-transformed afterwards. This gives the time series
$S[f,t]$ of power spectra $S(f)$. We have calculated the rms value
$\delta V^2_{rms} = \int \overline{S}(f)df$, the mean power
spectrum $\overline{S}(f)$ as a function of the acquisition time,
and the fractional histograms (i.e. the distribution of the power
contained in fixed frequency windows). We have also calculated the
second spectrum:
\begin{displaymath}
S_2(f,f_2) = \sum_{t=0}^{T_2} S[f,t] e^{-2i \pi f_2t},
\end{displaymath}
with $f_2$ the second order frequency. If the fluctuations are not
correlated on long time-scales (typically longer than 1s with our
resolution), $S_2(f,f_2)$ is expected to be frequency-independent for any $f$.
Conversely, if they exist, such correlations will lead to a $f_2$-dependency in the
second spectrum. We use this tool to estimate if any long
time-scale correlations exist in the fluctuations. It offers more
robustness than a direct correlation calculation for rather low
signal-to-noise ratios, which is our case here.

\section{Results and discussion}
To begin with, we briefly describe the noise for $I \gg I_c$, in the flux-flow regime.
The flux-flow regime is characterized by the linear relation $V = R_f (I-I_c)$, $R_f$ being the flux-flow resistance of the vortex lattice.
Since the flux-flow noise properties have been observed to be similar for all micro-bridges, regardless of their surface state and edges state, the samples will not be distinguished for the moment.
In general, the introduction of a periodic artificial
pinning potential leads to the observation of matching fields,
when the magnetic field matches exactly an integer number of
vortices between two pinning sites \cite{dots}. We have observed a
substantial increase of the critical current at matching fields in
the grooved sample \cite{scolathesis}. But, in order to avoid any
noise artifacts, the low dissipation requirement imposes to
perform measurements only at fields much higher than the matching
fields, where the critical current is moderate. These matching
effects can thus be ignored here.

\begin{figure}[htp!]
\begin{center}
\includegraphics[width=4.7cm,angle=270]{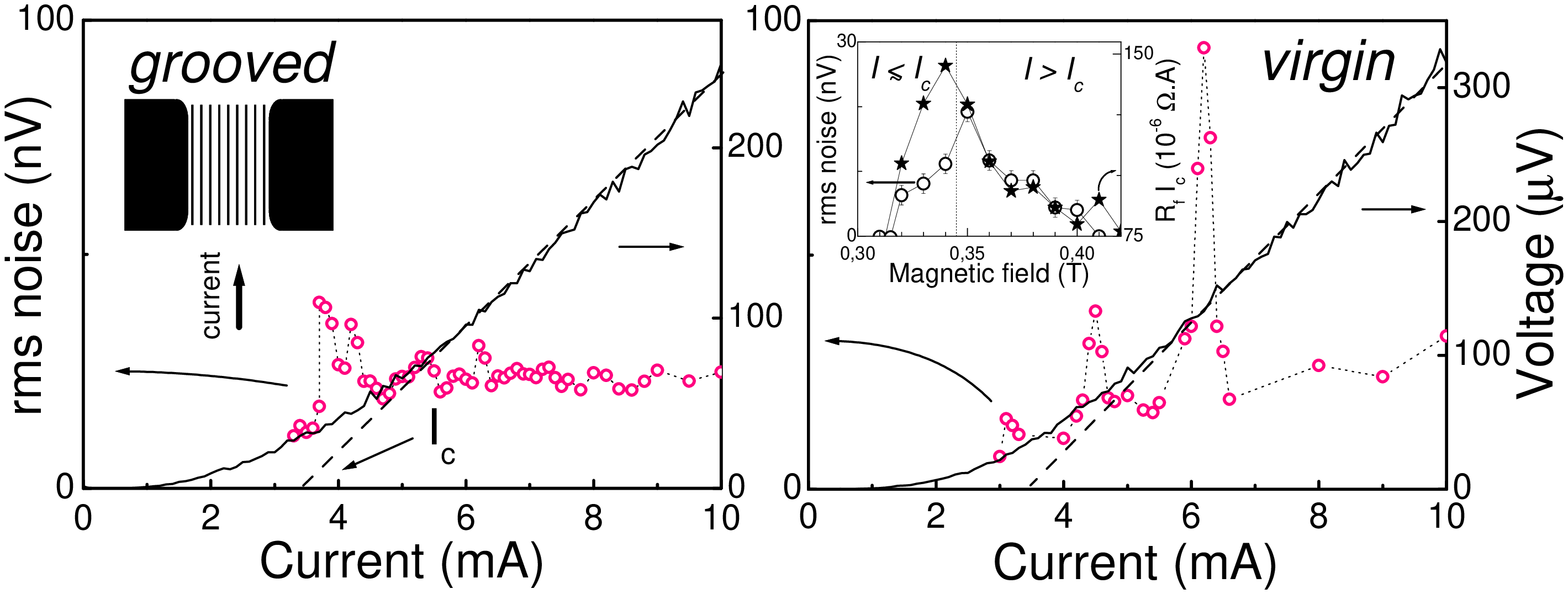}
\caption{Average voltage (solid line) along with the rms noise power $\delta V_{rms}$ ($\circ$) plotted against the driving current $I$, for the grooved and virgin bridges (the depinning noise points in the former are shown as an indication).
The linear extrapolation (dashed line) of the flux-flow voltage intercepts the $I$-axis at the overall critical current $I_c$.
The etched pattern on the micro-bridge is sketched on the left-hand graph.
Insert: magnetic field scan of the rms noise ($\circ$) at $I=7mA$, and the product $R_f\,I_c$ ($\star$).
On the right-hand side of the vertical line, the lattice is in flux-flow regime.
$T=4.2K,\, B = 0.35T$.}
\label{BruitdeI}
\end{center}
\end{figure}

\begin{figure}
\begin{center}
\includegraphics[width=4.7cm,angle=270]{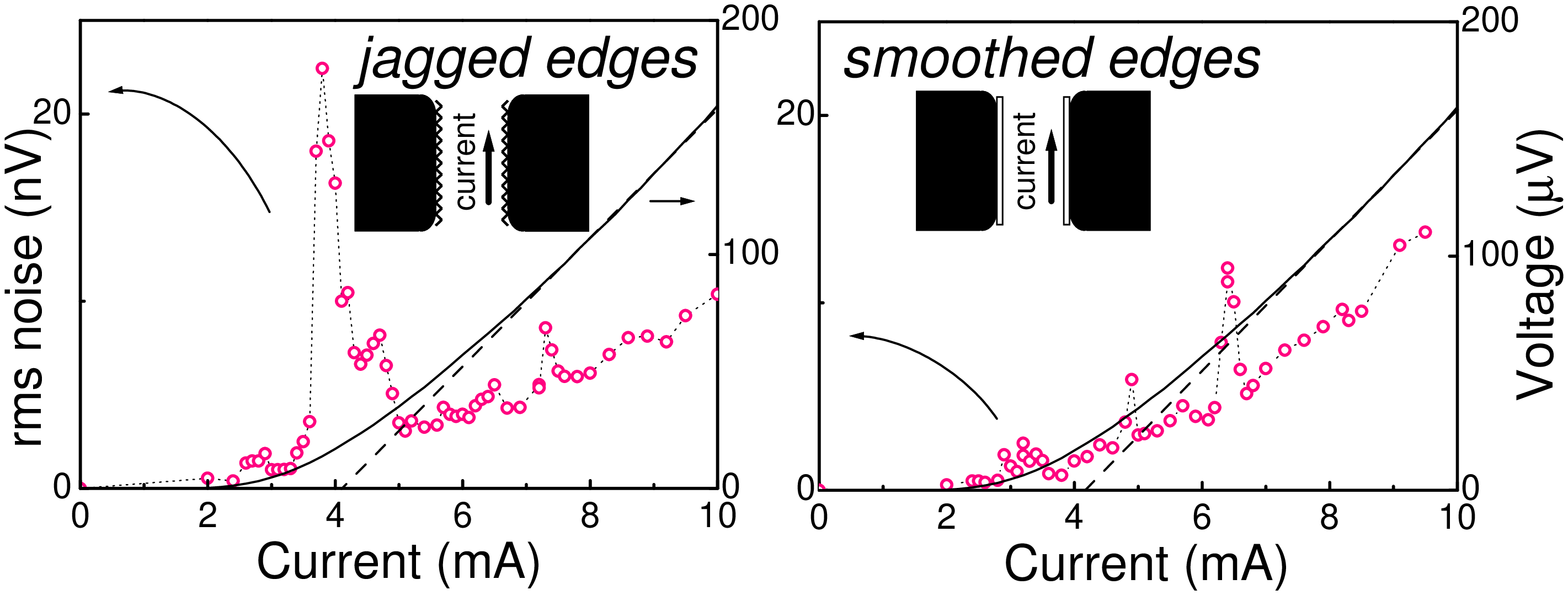}
\caption{Average voltage and rms noise versus the driving current, as in figure \ref{BruitdeI}.
The types of edges are sketched.
The etched pattern is made of two series of forty $200nm\times200nm$ squares (technical details can be found in ref.\cite{scolathesis}).
B = 0.32T.}
\label{edges}
\end{center}
\end{figure}

\begin{figure}
\centering{\resizebox{12.7cm}{!}{\includegraphics{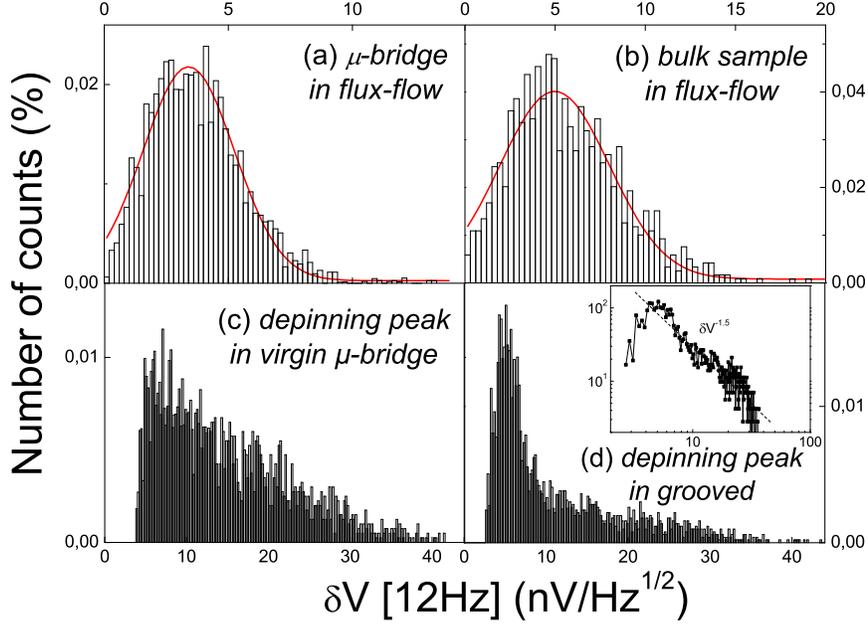}}}
\caption{Normalized histograms of the noise magnitude at $12 \pm
2Hz$ in micro-bridges (a,c,d, 7000 counts) and a bulk Nb sample (b, $12.10
\times 1.23 \times 0.22 mm^3$, 1200 counts) built from four
30 minute long series, for different driving conditions: (a)
$I = 7.5mA$; (b) $I = 4.4A \gg I_c = 1.61A$, $H = 0.28T$; (c) $I =
4.5mA$; (d) $I = 6mA$. $T = 4.2K$. Solid curves denote the
Gaussian fits.
Details on the measuring condition of the bulk sample can be found in \cite{scola06}.
Inset: log-log plot of the histogram (d).}
\label{histos}
\end{figure}

The flux-flow noise can be described as follows:
the rms value hardly depends on the driving current, and hence on the mean lattice velocity v$_L = E/B = \rho_f (J-J_c)/B$, (fig. \ref{BruitdeI} and \ref{edges}), and it varies like the product $R_f I_c$ (inset of fig. \ref{BruitdeI});
the noise spectrum ranges roughly below $1kHz$;
fractional histograms look clearly Gaussian (fig. \ref{histos}).
All those properties also depict the flux-flow noise in bulk samples \cite{placais94,scola05}.
These results are of importance, given that in a micro-bridge the vortex velocity v$_L$ can be much higher than in bulk samples, due to the largest values of the driving current density $(J-J_c)$.
Here, v$_L$ is typically three orders of magnitude higher than in bulk samples (up to $85m/s$ for $I = 10A$ compared to v$_L\lesssim 0.1m/s$ in bulk Nb).
This clearly shows that the vortex lattice velocity plays a minor role in the low frequency noise mechanism.
This contradicts the common descriptions of the noise coming from direct interactions of the vortices with some bulk pinning centers.
In such approaches, the noise should disappear at high speed, as the pinning interactions are expected to vanish (e.g. \cite{koshelev94,olson}).
In contrast, the observed properties support a description of the noise in terms of surface critical current fluctuations, which has been verified in bulk samples of low-$T_{c}$ superconductors \cite{placais94,scola06}.

We will now focus on the low current regime.
This corresponds to the non-linear part of the V(I) curve, where it is known that the vortex depinning is not homogeneous.
We observe that the rms noise exhibits peaks for some current values (fig. \ref{BruitdeI} and \ref{edges}) in most micro-bridges.
This excess noise can be associated to inhomogeneities of the flow before the flux-flow is reached.
Such inhomogeneities are likely to come from a non-uniform critical current.
To elucidate the origin of the peaks, micro-bridges were modified using a focused ion beam.
The edges of one micro-bridge were perfectly smoothed at the scale of the lattice, whereas the edges of another one were periodically jagged (see schematic drawing in figure \ref{edges}).
None of the critical current, the flux-flow resistivity, and the flux-flow noise behavior (weak monotonic dependency on $I$) were affected by the changes of the edges shape in our experimental conditions.
However, at low currents, the noise peaks are almost absent for the smoothed edges whereas a huge peak is visible for the jagged edges (figure \ref{edges}).
These results show that the depinning peaks are likely due to the irregularities of the edges.
Note that the Oersted field created by the transport current does vary along the jagged edges, but it is too small to explain our results: using the Biot-Savart formula, one estimates that the magnetic field at the edges induced by $10mA$ in our micro-bridges is only $10^{-3}T$, which is hundred times smaller than the external magnetic field.
Our interpretation of the phenomenon is that the edges irregularities cause a modulation of the width, and thereby of the critical current, along the micro-bridge.
The critical current is the lowest in the narrowest regions.
Conversely, the over-critical current, that is the driving force, is the highest in these locations.
Hence, the lattice in the narrowest regions is depinned for the lowest currents.
Shear areas are introduced and disappear gradually until the whole lattice flows.
It must be noted that the peaks always appear at the same current value, even if the current was increased from zero, or decreased from the flux-flow, or even after an over-critical temperature cycle.
This shows that these noise peaks are not history dependent, in agreement with their proposed topographical origin.

\begin{figure}
\centering{\resizebox{12.7cm}{!}{\includegraphics{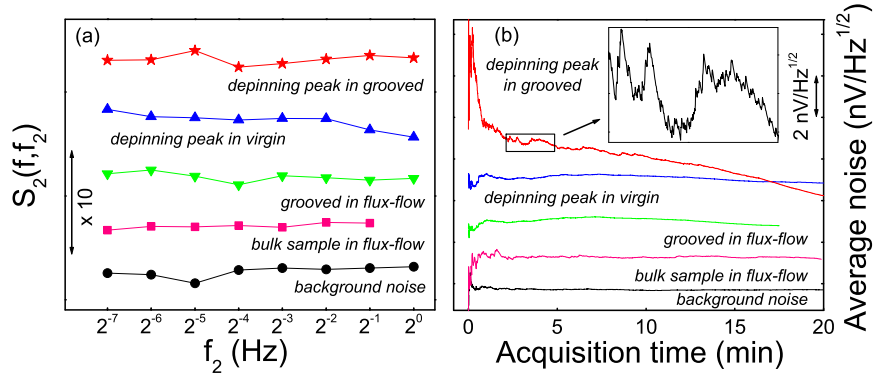}}}
\caption{(a): Second spectra of the noise power at 12$Hz$, represented by octaves in log-log scale.
(b): average noise at 12$Hz$ plotted against the acquisition duration;
an arbitrary offset is added for presentation purposes.
Both quantities are reported for different samples, dynamic regimes and surface states.
Calculations at 72$Hz$ yield similar results.}
\label{S2etStat}
\end{figure}

To go deeper in the understanding of these noise peaks, we
have investigated their statistics. The associated histograms are
clearly non-Gaussian (fig. \ref{histos}-c et d): the noise power
is distributed over a much larger range than in the flux-flow regime, while
the mean values are comparable. The noise power measured in the
grooved micro-bridge is the most unique: it exhibits roughly a power law
distribution, within the experimental range (inset of
\ref{histos}-d). To reveal the existence of long time correlations
between the fluctuations, the use of the second spectrum is an
efficient and complementary tool. The second spectra are
observed to be frequency-independent like in flux-flow, despite the presence of extreme values
in this regime (fig. \ref{S2etStat}-a): this
reveals the absence of long-time correlations (below 1$Hz$). This
means that when the applied current sets the lattice in a depinning peak, extreme
noise values are reached in an abrupt fashion, rather than a gradual one.
In summary, the fluctuations in the depinning peaks
can be characterized as erratic and intermittent, without memory.
Outside the peaks, the histograms asymmetry is less
pronounced, as though an intermediate regime between depinning and
flux-flow took place.
The two micro-bridges having their edges altered exhibit exactly the same statistic features as the virgin sample (non-Gaussian in the depinning peaks, when they exist, and gaussian in flux-flow).


The existence of non-Gaussian statistics during the depinning shows the difference of vortex flow between flux-flow and depinning.
Up to now, the different dynamic regimes of the vortex lattice have been investigated by means of numerical simulations (see e.g. \cite{olson}), and very little information have been collected from experiments.
Here, we bring an experimental evidence of the distinction between two dynamical regimes.
Our results can be interpreted in a similar way as the experiment carried out on NbSe$_2$.
Regardless of the discrepancies between Nb micro-bridges and
NbSe$_2$ samples, it seems that here again, the non-Gaussian
effects can be associated to spatial inhomogeneities of the
critical current \cite{marley95}. In our case, the fluctuations can not develop
over the whole lattice like in flux-flow, because they are
confined inside independent lattice slices. The non-Gaussian
effects are likely to be enhanced by the smallness of the micro-bridges
that limits the number of shear areas. Given the fact that a sum
of independent random variables tends to an approximate Gaussian
distribution from about five terms, we conclude that, following
our interpretation, only one or two shear areas take place
simultaneously, and the flux-flow is reached after a few depinning
steps. This sets an upper size of a few microns for one slice.

We now focus on the particular effects linked to the presence of the etched grooves.
Albeit the depinning noise regimes
with or without the grooves look qualitatively similar
(non-Gaussian power distributions and white second spectra, as
shown in fig. \ref{histos} and \ref{S2etStat}-a), the sample with
the grooves exhibits new and interesting features. In order to understand the differences, average noise powers,  at a given frequency, are
plotted as a function of the acquisition time $T_2$ under various
conditions (fig. \ref{S2etStat}-b). For stationary noise,
no time dependency is expected. This is clearly observed for all
samples in flux-flow, and also for the virgin sample in the low
current regime. But for the sample with the grooves, this low
current noise dramatically drifts with time.
This drift occurs at any observable frequency, at any current in the
depinning regime, and disappears in flux-flow.
It is important to note that
this non stationary noise appears only for bridges presenting a long power law tail in the noise power distribution.
This probability density decays as $1/x^{\alpha}$, with
$\alpha=1.5 < 2$. Such a slow decay magnifies the weight of rare events and the high values. In this case,
the statistics is not described by a Gaussian law, but by a
L\'{e}vy law, and the central limit theorem (CLT) does not apply
anymore \cite{levy}. As a consequence of the domination of a small
number of terms, the process ergodicity is lost. There is no
convergence towards any constant value, as compared to what
we had measured in the grooved sample during the depinning (fig.
\ref{S2etStat}-b inset). The dedicated term of such a random walk
with a long tail distribution is the L\'{e}vy flight. Levy
flights can be observed for some extreme cases of anomalous
diffusion in an Arrh\'{e}nius cascade. For such an activated
system in a tilted potential landscape, and in the limit where the
potential wells are high compared to the activation energy, the
trapping time does not converge towards any constant value
\cite{bardou99}. It is clear that in our case the diffusion of the
fluctuations are not thermally driven, as in a strict analogy with
the Arrh\'{e}nius cascade. The underlying instability process can
be here the local hang-and-release of vortices on the surface,
which was shown to be a potential candidate for the noise
mechanism \cite{placais94}.
Since the drift of the fluctuations are observed only for the
sample with the superficial grooves, this shows unambiguously that
they are very effective potential barriers for the vortex lattice
flow
\emph{fluctuations}. 
Violations of the CLT experimentally observed in physical systems, and
related to L\'{e}vy flights, have recently received an increasing
interest \cite{levy}.
These observations are often in direct
connection with an anomalous diffusion \cite{salomon93,latora}.
Thus, the return to the flux-flow regime, where the lattice flows homogeneously
at the scale of the system, is consistently marked by the return to stationary statistics (fig. \ref{S2etStat}-b).
From a practical point of view, such a L\'{e}vy-flight behavior implies also
that the size of the system and the number of measurements can
play an important role when measuring vortex lattice properties,
in some extreme, but experimentally accessible, cases.
More generally, by reducing the size of the system, we managed to observe a somewhat individual behavior, as opposed to the averaged one observable in large systems where the CLT is valid.
Our results underline how crucial fluctuations analysis can be for studying physical phenomena at small scale.

To conclude, we have reported a detailed voltage noise
investigation in superconducting Nb micro-bridges. Clear
experimental signatures distinguished the flux-flow regime at high
currents and the depinning regime at low currents. The former is always Gaussian
and stationary, which is quite comparable to common observations
made in bulk samples. In contrast, the latter is characterized by
non-Gaussian effects observable in the presence of spatial
variations of the critical current. In addition to these
non-Gaussian effects, an artificial surface pinning potential
behaves like high potential barriers, and induces an abnormal
diffusion of the fluctuations. This is, to our knowledge, the first
experimental observation of a L\'{e}vy flight-like behavior
of a vortex lattice.
This experiment is an example of the new statistical behavior that can arise from a drastic reduction of the size of the system.

\section*{Acknowledgement}
This work is supported by `` La R\'{e}gion Basse-Normandie ''.


\end{document}